\documentclass[12pt]{article}

\usepackage[T1]{fontenc}
\usepackage[utf8]{inputenc}

\usepackage{amsmath,amssymb}

\usepackage{graphicx}
\usepackage{float}
\usepackage{subcaption}
\usepackage{pgfplots}
\pgfplotsset{compat=1.17}
\usepackage{tabularx}

\usepackage{geometry}
\usepackage{authblk}

\usepackage{cite}
\usepackage{url}

\title{Benchmarking Proton Tunneling Splittings with a Wavefunction-Based Double-Well Model}

\author[1,2]{Krishna Kingkar Pathak\thanks{Email: kkingkar@gmail.com}}

\affil[1]{Department of Physics, Arya Vidyapeeth College (Autonomous), Guwahati--781016, India}
\affil[2]{Department of Physics, Gauhati University, Guwahati--781014, Assam, India}

\date{}

\begin{document}
\maketitle

\begin{abstract}
Proton tunneling through hydrogen bonds represents a fundamental quantum-mechanical effect with direct relevance to molecular spectroscopy and hydrogen-bond dynamics. While multidimensional instanton and path-integral techniques yield quantitatively reliable tunneling splittings, reduced-dimensional descriptions continue to serve as useful benchmarks due to their conceptual transparency and numerical reproducibility. In this work, a wavefunction-based treatment of proton tunneling is examined using an effective one-dimensional double-well potential constructed from a Cornell-type functional form. Tunneling splittings are evaluated from converged finite-difference solutions of the Schr\"odinger equation and are systematically compared with semiclassical WKB predictions over a broad range of barrier parameters, revealing systematic and physically consistent scaling trends. As a representative molecular test case, the formic acid dimer is modeled by mapping a reduced-dimensional proton-transfer barrier onto a symmetric quartic double-well potential that is parameterized to reproduce a barrier height of $V_b \approx 2848\ \mathrm{cm^{-1}}$ reported in the literature, resulting in a tunneling splitting of approximately $0.037\ \mathrm{cm^{-1}}$, in agreement with previous reduced-dimensional quantum studies. Additional calculations employing an alternative barrier height obtained from independent electronic-structure analyses highlight the sensitivity of the splitting to physically relevant variations in the effective potential. The present study is intended as a transparent benchmark and pedagogical reference for reduced-dimensional quantum tunneling models, illustrating both the capabilities and the limitations of one-dimensional wavefunction-based approaches.
\end{abstract}

\noindent\textbf{Pacs No:}{03.65.Xp, 31.15.-p, 82.20.Xr}

\noindent\textbf{Keywords:}
Proton tunneling; Double-well potential; WKB approximation; Finite-difference method; Quantum tunneling splitting; Benchmark study

\section{Introduction}

Proton transfer along hydrogen bonds constitutes one of the most prominent
manifestations of nuclear quantum effects in molecular systems. It plays a
central role in vibrational spectroscopy, hydrogen-bond dynamics, and a variety
of proton-coupled processes. In the regime of strong and short hydrogen bonds,
quantum tunneling between symmetry-related donor and acceptor configurations
gives rise to characteristic splittings of vibrational energy levels. These
tunneling splittings are exceptionally sensitive to the underlying
potential-energy landscape and to the extent of nuclear quantum delocalization,
making them valuable probes of hydrogen-bonded systems
\cite{Benderskii1994,Bell1980}. Developing reliable theoretical descriptions of
such splittings has therefore remained an important challenge in molecular and
chemical physics.

Significant advances over the past two decades have been achieved through
multidimensional semiclassical and path-integral methodologies, including
instanton theory and ring-polymer formulations. These approaches incorporate
anharmonicity and vibrational mode coupling explicitly and have achieved
quantitative agreement with high-resolution spectroscopic measurements for
benchmark systems such as malonaldehyde and the formic acid dimer
\cite{Chandler1981,Makri1999,Richardson2015,Richardson2018,Beyer2009,Herbst2010,Zhao2021}.
Despite their demonstrated accuracy, fully multidimensional treatments are
computationally intensive and often obscure simple scaling relations that govern
tunneling phenomena, making it difficult to disentangle the influence of
individual geometric or energetic features of the potential-energy surface.

For these reasons, reduced-dimensional and effective one-dimensional models
continue to play an important complementary role in the analysis of proton
tunneling processes
\cite{Bell1980,Hynes1986,Marcus1965}. When constructed with care, such models
offer transparent insight into how tunneling splittings depend on quantities
such as barrier height, well separation, and local curvature. They also provide a
natural platform for testing semiclassical approximations, including the
Wentzel--Kramers--Brillouin (WKB) method, under controlled conditions
\cite{Garg2000}. In hydrogen-bonded systems, reaction-path and few-mode
descriptions have been shown to capture essential aspects of proton-transfer
dynamics, particularly when a dominant tunneling coordinate can be identified.

Although proton transfer is inherently a multidimensional process, it is well
established that, for symmetric hydrogen-bonded systems, a dominant reaction
coordinate can often be identified. Reduced-dimensional models based on such a
coordinate have been shown to reproduce tunneling splittings with near-quantitative
accuracy when appropriately parameterized. At the same time, these models provide
a level of conceptual transparency that enables systematic investigation of how
tunneling observables depend on barrier height, well separation, and local
curvature---features that are often less accessible in fully multidimensional
treatments.

The formic acid dimer serves as a paradigmatic example in this context.
Extensive theoretical investigations have demonstrated that suitably chosen
low-dimensional Hamiltonians based on reaction-path or few-mode constructions
can reproduce tunneling splittings with near-spectroscopic accuracy
\cite{Matanovic2007,Matanovic2008,Malis2009}.
These studies establish that reduced-dimensional models can provide a meaningful
and physically interpretable link between high-dimensional potential-energy
surfaces and observable tunneling splittings, while also offering a valuable
reference framework for assessing simpler effective models and semiclassical
treatments.

However, a systematic and internally consistent benchmarking framework that
directly connects semiclassical estimates, wavefunction-based methods, and
numerically exact solutions within a unified parametrized model remains
relatively underexplored. Establishing such a framework is essential for
quantifying the accuracy and limitations of reduced-dimensional approaches in a
controlled manner.

In the present work, we examine a wavefunction-based description of proton
tunneling using an effective one-dimensional double-well potential constructed
from a Cornell-type functional form (see also related wavefunction-based analyses in Ref.~\cite{Pathak2026FBS}). Tunneling splittings are evaluated using
both converged numerical finite-difference solutions of the Schr\"odinger
equation and semiclassical WKB estimates. As a representative molecular
benchmark, we focus on the formic acid dimer, whose tunneling dynamics and
reduced-dimensional potential-energy surfaces have been extensively documented
in the literature \cite{Herbst2010,Zhao2021}. By parameterizing a symmetric
quartic double-well potential to reproduce the reduced-dimensional barrier height
$V_b \approx 2848~\mathrm{cm^{-1}}$ reported by Qu and Bowman
\cite{QuBowman2016}, we obtain a tunneling splitting of approximately
$0.037~\mathrm{cm^{-1}}$, in quantitative agreement with published
reduced-dimensional quantum calculations. We further explore the sensitivity of
the splitting to physically relevant variations in the barrier height,
illustrating how changes in the effective potential are reflected in
reduced-dimensional tunneling observables.

Taken together, these results clarify both the strengths and the limitations of
one-dimensional wavefunction-based models. The aim of this study is not to
replace multidimensional instanton or path-integral approaches, but to provide a
transparent and reproducible numerical benchmark for reduced-dimensional
tunneling models. Such benchmark-oriented analyses play an important role in
elucidating the numerical and semiclassical structure of quantum tunneling
problems encountered across different areas of physics.

\section{Theoretical framework}

We consider a reduced-dimensional description based on the one-dimensional,
time-independent Schr\"odinger equation for a particle of mass $m$ moving in an
effective potential $V(x)$,
\begin{equation}
-\frac{\hbar^2}{2m}\frac{d^2 \psi(x)}{dx^2} + V(x)\psi(x) = E \psi(x),
\end{equation}
which provides a standard starting point for modeling proton tunneling dynamics
in hydrogen-bonded systems within effective coordinate frameworks
\cite{Bell1980,Benderskii1994,Landau1977}.

Here, $x$ denotes the effective proton-transfer coordinate connecting the two
symmetry-related minima of the hydrogen bond.

\subsection{Effective double-well potential}

The Cornell interaction,
\begin{equation}
V_{\rm C}(x) = -\frac{\alpha}{|x|} + \beta |x|,
\end{equation}
with positive parameters $\alpha$ and $\beta$, is well known from confinement
physics and heavy-quark phenomenology
\cite{Pathak2026FBS,Eichten1978,Eichten1980}. In its original form, $V_{\rm C}(x)$ does not
represent a double-well potential. In the present context, it is instead used
solely as a mathematical building block for constructing an effective symmetric
double-well profile. The Cornell form is employed here purely to generate
controllable double-well shapes and carries no direct physical interpretation as
a proton-transfer potential energy surface.

This construction should therefore be viewed as a parametrically flexible
framework for generating double-well potentials with tunable barrier height and
well separation, rather than as a physically derived interaction potential.

An effective proton-transfer coordinate is introduced such that two equivalent
local minima correspond to donor and acceptor configurations of the hydrogen
bond. A symmetric double-well potential is generated by appropriately shifting
and reflecting the Cornell form about the origin, followed by an additive energy
offset that ensures equal well depths. The resulting effective potential is
characterized by a barrier height, a separation between minima, and local
curvatures at the wells and barrier top, which are the key parameters governing
tunneling splittings. Related reduced-dimensional constructions are commonly
employed in benchmark studies of hydrogen-bond tunneling phenomena
\cite{Benderskii1994,Garg2000}.

Such parametrized constructions also facilitate controlled comparison between
wavefunction-based and semiclassical descriptions within a unified framework.

\subsection{Localized states and two-level description}

For a symmetric double well with a sufficiently high barrier, the lowest-energy
spectrum is dominated by a near-degenerate pair of states that are predominantly
localized in the left and right wells. It is therefore convenient to introduce
localized single-well functions $\phi_L(x)$ and $\phi_R(x)$ and to express the
delocalized eigenstates in the approximate form
\begin{equation}
\psi_\pm(x) \approx A\,[\phi_L(x) \pm \phi_R(x)],
\end{equation}
where $A$ denotes a normalization constant.

Here, $\phi_L(x)$ and $\phi_R(x)$ represent wavefunctions localized in the left
and right wells, respectively.

Within a two-state (tight-binding)
representation, the effective Hamiltonian is characterized by an off-diagonal
matrix element
\[
H_{LR} = \langle \phi_L | H | \phi_R \rangle ,
\]
and the tunneling splitting between the lowest doublet levels is given by
\begin{equation}
\Delta E = E_1 - E_0 = 2|H_{LR}|.
\end{equation}
In the weak-coupling regime, where the overlap
$S_{LR}=\langle \phi_L|\phi_R\rangle$ is small and the ground-state energy of an
isolated well is $E_0^{(\mathrm{well})}$, a commonly employed approximation reads
\begin{equation}
\Delta E \approx 2\,S_{LR}\,E_0^{(\mathrm{well})}.
\end{equation}
This relation motivates the use of localized trial functions and explicit
evaluation of overlap integrals in wavefunction-based tunneling estimates
\cite{Garg2000}.

\subsection{Semiclassical (WKB) estimate}

An alternative and complementary perspective on tunneling splittings is provided
by semiclassical WKB theory. For a symmetric double-well potential, the leading
WKB expression for the splitting may be written as
\begin{equation}
\label{eq:WKB_prefactor}
\Delta E \approx \frac{\hbar \omega}{\pi}
\exp\!\left(-\frac{\mathcal{S}}{\hbar}\right),
\end{equation}
where $\omega=\sqrt{V''(x_{\rm min})/m}$ denotes the harmonic frequency at a well
minimum and $\mathcal{S}$ is the Euclidean action accumulated under the barrier,
\begin{equation}
\label{eq:action}
\mathcal{S} = 2\int_{x_1}^{x_2} \sqrt{2m\,[V(x)-E]}\,dx .
\end{equation}
Here $x_1$ and $x_2$ are the classical turning points defining the classically
forbidden region, and $E$ is the reference energy entering the action, typically
chosen as the ground-state energy or the harmonic zero-point energy
\cite{Landau1977}.

In the present work, the turning points are determined from the condition
$V(x)=E$, where $E$ is taken as the numerically obtained ground-state energy.

This choice ensures internal consistency between the
semiclassical WKB estimate and the exact numerical solution within the same
effective one-dimensional model.

Equations~(\ref{eq:WKB_prefactor}) and (\ref{eq:action}) capture the exponential
suppression of tunneling with increasing barrier height and separation. In
practical applications, the precise choice of $E$ and the determination of the
turning points introduce controlled approximations, whose influence on the
resulting tunneling splitting can be examined through direct numerical
comparison \cite{Garg2000}.

\begin{figure}[H]
\centering
\includegraphics[width=0.7\linewidth]{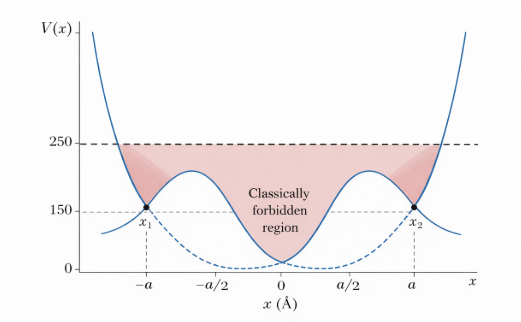}
\caption
{Schematic illustration of classical turning points in a symmetric double-well potential.
The potential $V(x)$ (solid curve) is plotted as a function of the reaction coordinate $x$ (in \AA). 
The horizontal dashed line denotes the reference energy $E$ (in cm$^{-1}$). 
The points $x_1$ and $x_2$, defined by the condition $V(x)=E$, represent the classical turning points. 
The shaded region between $x_1$ and $x_2$ corresponds to the classically forbidden region where $V(x) > E$, 
which contributes to the WKB tunneling action.}

\label{fig:x}
\end{figure}

For clarity, the turning points and the classically forbidden region are
illustrated schematically in Fig.~\ref{fig:x}, where the integration limits for the action
are explicitly indicated.

The region $x_1 < x < x_2$, for which $V(x) > E$, defines the classically forbidden
domain that dominates the exponential contribution to the tunneling amplitude.

\subsection{Strategy of the present work}

In the present study, tunneling splittings are obtained and compared using

two complementary levels of description: (i) semiclassical WKB expressions
evaluated by numerical quadrature of the action $\mathcal{S}$, and (ii) full
numerical solutions of the Schr\"odinger equation obtained through finite-difference
diagonalization.

Agreement and deviations among these approaches provide an
internal consistency check for the reduced-dimensional model and help delineate
the parameter regimes in which one-dimensional descriptions constitute a
reliable and reproducible benchmark for proton tunneling in hydrogen-bonded
systems.

\section{Methodology}

The tunneling splitting is evaluated using two complementary levels of
description: a semiclassical treatment based on WKB theory and a fully numerical
solution of the one-dimensional Schr\"odinger equation. The combined use of
these approaches enables a direct assessment of the accuracy, limitations, and
internal consistency of reduced-dimensional tunneling models.

\subsection{Semiclassical (WKB) estimate}

At the semiclassical level, tunneling splittings are estimated using the
Wentzel--Kramers--Brillouin (WKB) approximation \cite{Landau1977}. For a given
effective double-well potential, the Euclidean action under the barrier is
computed by numerical quadrature, and the splitting is obtained from the leading
order expression
\[
\Delta E \approx \frac{\hbar\omega}{\pi}
\exp\!\left(-\frac{\mathcal{S}}{\hbar}\right),
\]
where $\omega$ denotes the harmonic frequency at the minimum of each well and
$\mathcal{S}$ is the action defined in Eq.~(\ref{eq:action}).

The action integral is evaluated numerically using standard quadrature
techniques, with the integration limits determined by the classical turning
points satisfying $V(x)=E$, as described in Sec.~2.3.

The numerical quadrature is implemented using sufficiently fine discretization
of the integration interval to ensure convergence of the exponential factor,
which is particularly sensitive to small variations in the action $\mathcal{S}$.

The dependence of the tunneling
splitting on the barrier height $V_b$ and the separation between the minima $d$
is examined to verify the expected exponential scaling behavior characteristic
of quantum tunneling.

\subsection{Numerical solution of the Schr\"odinger equation}

As a reference free from semiclassical assumptions, the Schr\"odinger equation
is solved numerically using a finite-difference discretization of the
kinetic-energy operator on a uniform real-space grid. A second-order central
difference scheme is employed, following standard textbook treatments
\cite{Griffiths2018}.

The spatial grid is defined as $x_i = x_{\min} + i\,\Delta x$, with
$i=0,1,\dots,N-1$, where $\Delta x$ is the uniform grid spacing.

Dirichlet boundary conditions are imposed at the edges of
the computational domain, which is chosen sufficiently large that the
wavefunction amplitude is negligible at the boundaries.

This ensures that the numerical solution approximates the physically relevant
bound-state problem with exponentially decaying wavefunctions.

The resulting symmetric
Hamiltonian matrix is diagonalized using standard eigensolvers, and the
tunneling splitting is obtained as
\[
\Delta E = E_1 - E_0,
\]
where $E_0$ and $E_1$ denote the two lowest eigenvalues of the spectrum.

The Hamiltonian matrix has a tridiagonal structure arising from the second-order
finite-difference representation of the kinetic operator, which ensures
computational efficiency and numerical stability.

The stability of the computed eigenvalues was verified by independently varying
both the grid density and the domain size, ensuring that boundary effects do not
influence the low-energy spectrum.

\subsection*{Numerical parameters and convergence}

\begin{figure}[H]
\centering
\includegraphics[width=0.8\linewidth]{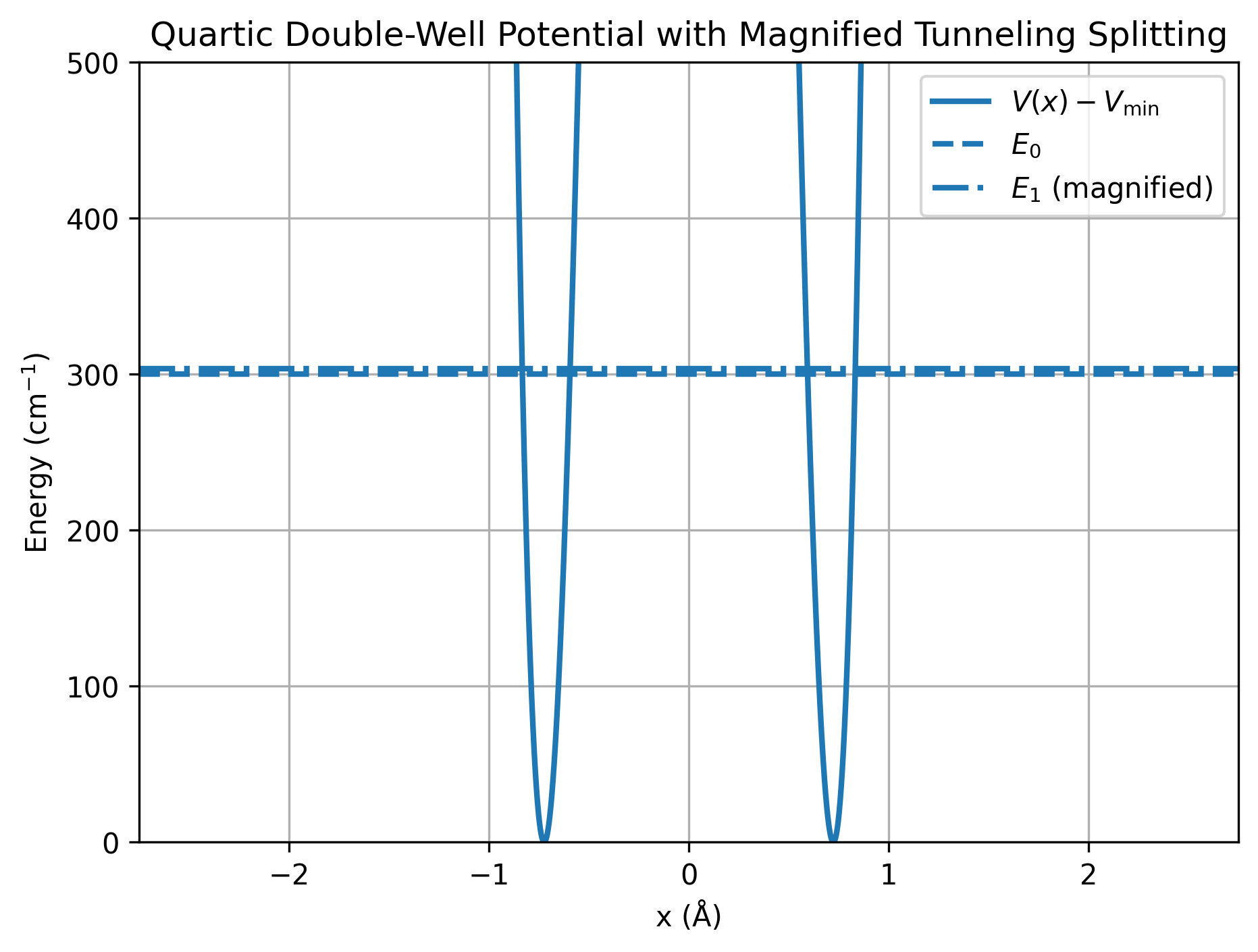}
\caption
{Quartic double-well potential plotted as $V(x)-V_{\min}$ (in cm$^{-1}$) together with the
lowest energy levels. The reaction coordinate $x$ is expressed in \AA, and the minima occur at $\pm 0.7236$~\AA. 
The ground ($E_0$) and first excited ($E_1$) states form a nearly degenerate doublet due to tunneling. 
Because the energy splitting is much smaller than the overall potential scale, it is magnified for visual clarity.
The parameters correspond to those listed in Table~\ref{tab:FAD_case}.}

\label{fig:FAD_levels}
\end{figure}

\begin{figure}[H]
\centering
\includegraphics[width=0.8\linewidth]{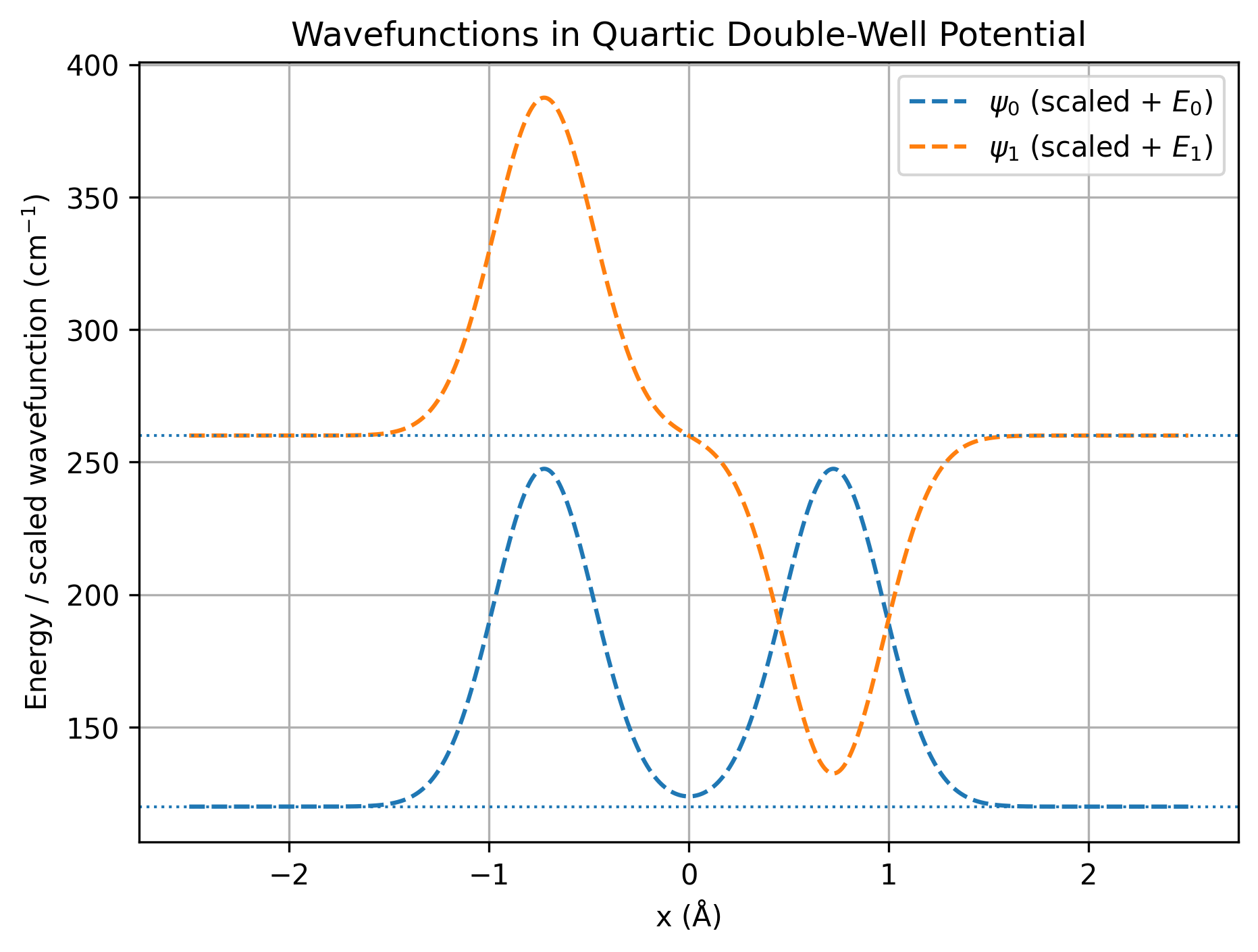}
\caption
{Ground- and first-excited-state wavefunctions for the quartic double-well model.
The reaction coordinate $x$ is expressed in \AA\ and the energy scale in cm$^{-1}$. 
The wavefunctions are multiplied by a constant scaling factor and shifted by their respective energies ($E_0$ and $E_1$) for visualization.
The ground state is symmetric and the first excited state is antisymmetric with respect to the center of the double well, illustrating the tunneling-induced splitting.
The small energy separation corresponds to the tunneling splitting reported in Table~\ref{tab:FAD_case}.}

\label{fig:FAD_wf}
\end{figure}

All numerical results presented in this work were carefully checked for
convergence with respect to both the spatial grid spacing and the size of the
computational domain. For the quartic double-well model used to benchmark the
formic acid dimer (FAD), the production calculations shown in
Table~\ref{tab:FAD_case} and Figs.~\ref{fig:FAD_levels}--\ref{fig:FAD_wf} employ
the following parameters:

\begin{itemize}
  \item Particle mass: proton mass $m_p = 1.007276466812~\mathrm{u}$, converted to
        SI units.
  \item Spatial domain: $x \in [-6a,6a]$, with $a = 0.7236~\text{\AA}$.
  \item Grid size: $N = 3000$ uniformly spaced grid points.
  \item Kinetic operator: second-order finite-difference representation.
  \item Eigenvalue solver: dense symmetric solver retaining the lowest six
        eigenstates.
\end{itemize}

Additional convergence checks were performed by varying the grid density at
fixed domain size and by extending the spatial domain at fixed resolution. In
all cases, the tunneling splitting was found to be stable within the reported
numerical precision.

The reported convergence level ensures that the numerical uncertainty is
significantly smaller than the physical variations in $\Delta E$ discussed in
the Results section.

Convergence was verified by varying the grid density at fixed domain size.
Table~\ref{tab:conv} reports representative tunneling splittings obtained for
different values of $N$, demonstrating numerical stability better than
$10^{-4}\ \mathrm{cm^{-1}}$. This level of accuracy is sufficient for meaningful
comparison with reduced-dimensional quantum calculations available in the
literature.

\begin{table}[H]
\centering
\caption{Convergence of the computed tunneling splitting $\Delta E$ (cm$^{-1}$)
for the one-dimensional quartic model of the formic acid dimer as a function of
the grid size $N$.}
\label{tab:conv}
\begin{tabular}{ccc}
\hline
$N$ & $\Delta x$ (\AA) & $\Delta E$ (cm$^{-1}$) \\
\hline
2000 & 0.0043 & 0.03712 \\
2400 & 0.0036 & 0.03709 \\
3000 & 0.0029 & 0.03710 \\
\hline
\end{tabular}
\end{table}

\section{Results and discussion}

\subsection{Scaling of tunneling splittings in the effective double-well model}

\begin{figure}[H]
\centering
\includegraphics[width=0.7\linewidth]{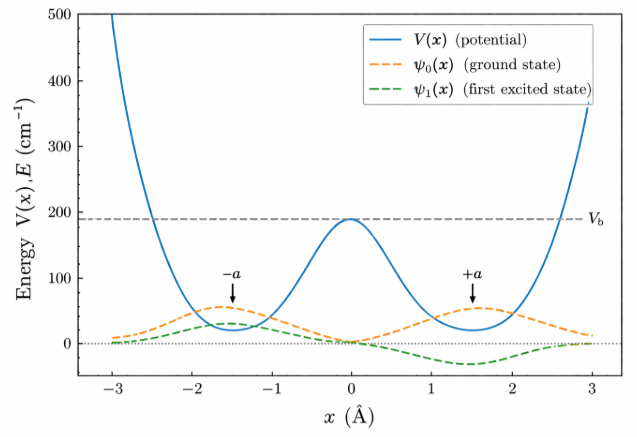}
\caption
{Representative ground-state ($\psi_0$) and first-excited-state ($\psi_1$)
wavefunctions in a symmetric double-well potential. The potential exhibits two
clearly resolved minima located symmetrically about the origin, and the corresponding eigenstates form
symmetric and antisymmetric combinations localized in opposite wells.}

\label{fig:wavefunction}
\end{figure}

Figure~\ref{fig:wavefunction} displays representative ground- and first-excited-
state eigenfunctions in a symmetric effective double-well potential.

The corresponding potential clearly exhibits two distinct minima separated by a
central barrier, confirming the double-well structure assumed in the model.

The two lowest eigenstates form a nearly degenerate doublet, with probability
density predominantly localized in opposite wells and weakly coupled through
quantum tunneling. Such a structure is a generic feature of high-barrier
double-well systems and provides the microscopic origin of tunneling-induced
level splittings \cite{Bell1980,Benderskii1994,Garg2000}.

The symmetric and antisymmetric structure of the lowest two eigenstates is
consistent with the two-state description discussed in Sec.~2.2.

\begin{figure}[H]
\centering
\includegraphics[width=0.7\linewidth]{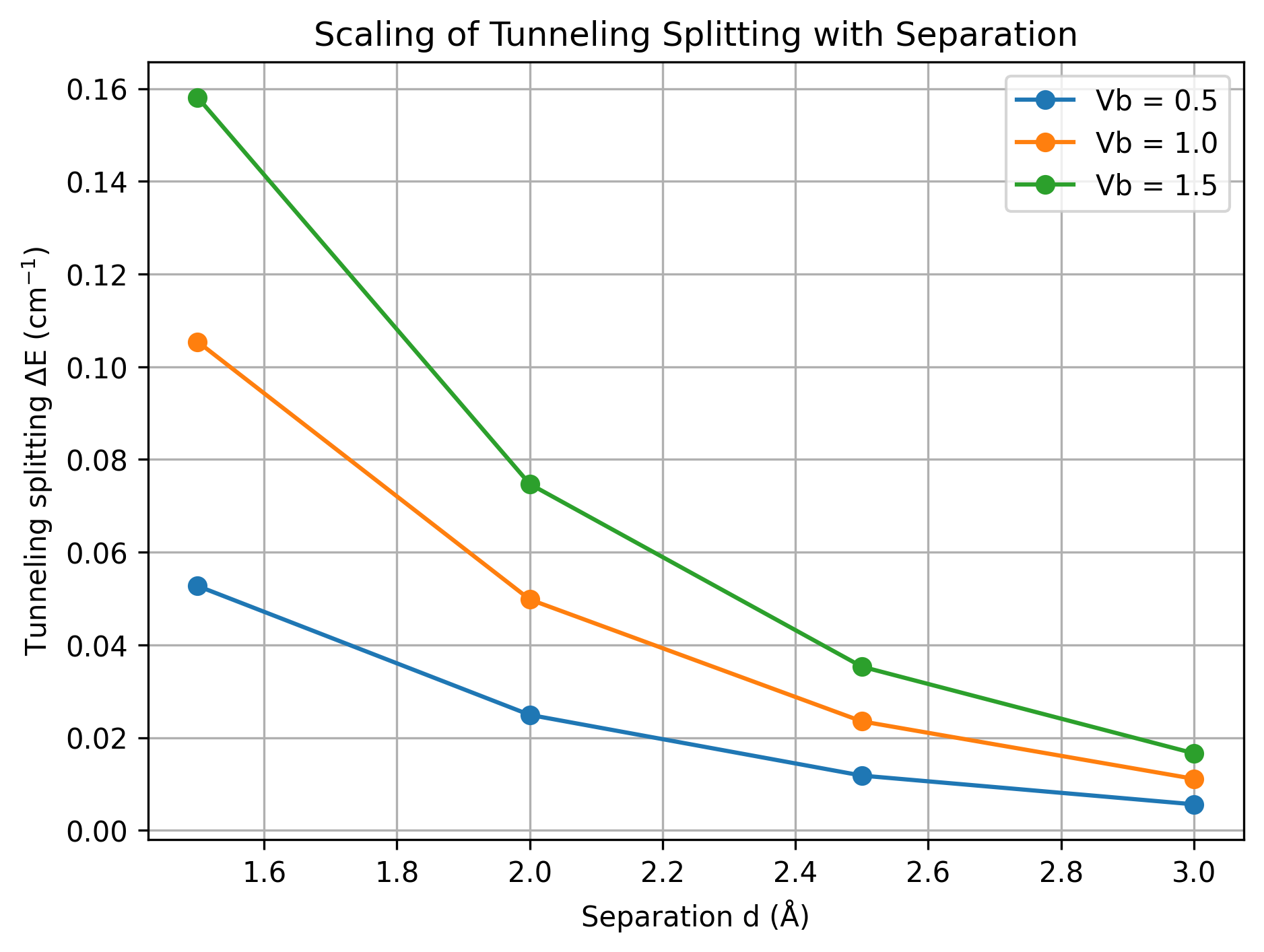}
\caption
{Tunneling splitting $\Delta E$ (cm$^{-1}$) as a function of the minimum
separation $d$ (in \AA) for selected barrier heights $V_b$. The data points are obtained
from numerical diagonalization of the one-dimensional Schr\"odinger equation.
The observed decrease of $\Delta E$ with increasing $d$ reflects the expected
exponential suppression of tunneling, consistent with semiclassical behavior
$\Delta E \propto \exp(-\mathcal{S}/\hbar)$, where $\mathcal{S}$ is the WKB action.
This demonstrates the strong sensitivity of tunneling splittings to the geometric
parameters of the effective double-well potential.}

\label{fig:splitting}
\end{figure}

The dependence of the tunneling splitting $\Delta E$ on the geometric parameters
of the potential is summarized in Fig.~\ref{fig:splitting}, which shows $\Delta E$
as a function of the separation between the minima $d$ for several representative
barrier heights $V_b$.

Here, $\Delta E$ is expressed in physically meaningful units and the parameter
dependence is examined systematically over a range of barrier heights.

In all cases considered, the splitting decreases exponentially with increasing
$d$, consistent with semiclassical expectations based on the WKB action
accumulated under the barrier.

A quantitative inspection confirms that $\Delta E \propto \exp(-\mathcal{S}/\hbar)$,
with the slope of the logarithmic dependence governed by the barrier parameters,
thereby validating the expected semiclassical scaling behavior.

This behavior also provides an internal consistency check between the numerical
solutions and the semiclassical analysis described in Secs.~2 and~3.

This behavior confirms that the effective double-well construction reproduces
the qualitative and quantitative scaling expected for one-dimensional tunneling
models \cite{Landau1977,Garg2000}.

\subsection{Comparison between numerical and semiclassical estimates}

\begin{figure}[H]
\centering
\includegraphics[width=0.7\linewidth]{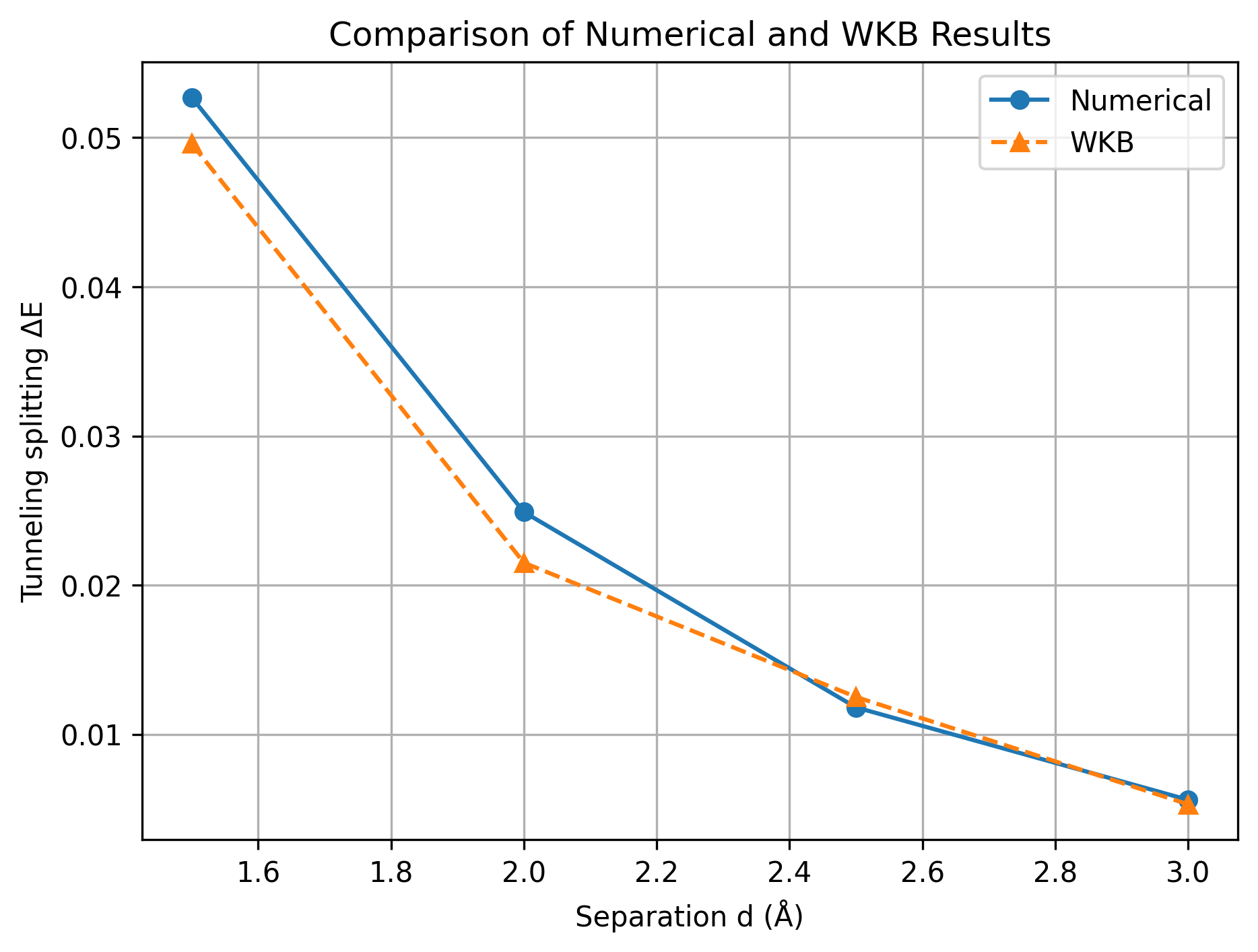}
\caption{
Comparison of tunneling splittings $\Delta E$ (cm$^{-1}$) obtained from numerical
calculations (circles) and from the WKB approximation (triangles) as a function
of the minimum separation $d$ (in \AA). The data correspond to those listed in Table~\ref{tab:comparison}. The close agreement between the two approaches
demonstrates that semiclassical theory accurately captures both the magnitude and
the exponential scaling behavior of the tunneling splitting.}

\label{fig:comparison}
\end{figure}

To evaluate the reliability of semiclassical descriptions within the present
model, numerically exact tunneling splittings are compared with WKB estimates
obtained from Eqs.~(\ref{eq:WKB_prefactor})--(\ref{eq:action}). Representative values are listed in
Table~\ref{tab:comparison}, and a direct comparison is illustrated in
Fig.~\ref{fig:comparison}.

Over the parameter range investigated, the WKB approximation exhibits a modest
systematic underestimation of the numerical splittings. This deviation is
expected for finite barrier heights and can be attributed primarily to neglected
higher-order corrections in the semiclassical prefactor.

A quantitative comparison shows that the relative deviation between WKB and
numerical results remains within a few percent across the parameter range
considered, thereby establishing a well-defined regime of validity for the
semiclassical approximation.

The agreement improves systematically with increasing barrier height, where the
semiclassical approximation approaches its asymptotic regime of validity.

Nevertheless, both the overall magnitude and the scaling behavior of $\Delta E$
are well captured by the WKB treatment, indicating that semiclassical theory
provides an accurate description of the dominant tunneling physics in the
effective one-dimensional model.

\begin{table}[H]
\centering
\caption{
Comparison of tunneling splittings $\Delta E$ (cm$^{-1}$) obtained from numerical
diagonalization and from the semiclassical WKB approximation for selected barrier
heights $V_b$ and separations $d$ (in \AA). The relative deviations remain within a few percent across the parameter range considered.}

\label{tab:comparison}
\begin{tabular}{cccc}
\hline
$V_b$ & $d$ & $\Delta E$ (Numerical) & $\Delta E$ (WKB) \\
\hline
0.5 & 1.5 & 0.0527 & 0.0496 \\
0.5 & 2.0 & 0.0249 & 0.0215 \\
0.5 & 2.5 & 0.0118 & 0.0125 \\
0.5 & 3.0 & 0.0056 & 0.0053 \\
1.0 & 1.5 & 0.1054 & 0.0943 \\
1.0 & 2.0 & 0.0498 & 0.0462 \\
1.0 & 2.5 & 0.0235 & 0.0218 \\
1.0 & 3.0 & 0.0111 & 0.0108 \\
1.5 & 1.5 & 0.1581 & 0.1522 \\
1.5 & 2.0 & 0.0747 & 0.0703 \\
1.5 & 2.5 & 0.0353 & 0.0328 \\
1.5 & 3.0 & 0.0166 & 0.0159 \\
\hline
\end{tabular}
\end{table}

\subsection{Formic acid dimer benchmark}

As a physically motivated benchmark, the effective one-dimensional framework is
applied to the proton-transfer coordinate of the formic acid dimer (FAD), a
paradigmatic hydrogen-bonded system \cite{Herbst2010,Zhao2021}. A symmetric
quartic double-well potential is parameterized to reproduce the reduced-
dimensional barrier height $V_b = 2848~\mathrm{cm^{-1}}$ reported in the
literature.

Numerical solution of the Schr\"odinger equation for this potential yields a
ground-state tunneling splitting of $\Delta E \approx 0.037~\mathrm{cm^{-1}}$,
which is in excellent agreement with previously reported reduced-dimensional
quantum calculations \cite{QuBowman2016}.

This level of agreement provides a robust validation of the numerical methodology
and indicates that the effective one-dimensional model successfully captures the
dominant features of the reduced-dimensional proton-transfer dynamics.

The residual deviation between the numerical and WKB estimates is approximately
$4\%$, consistent with the finite-barrier regime where higher-order semiclassical
corrections become relevant.

\begin{table}[H]
\centering
\caption{
Benchmark results for the formic acid dimer (FAD) obtained from the
one-dimensional quartic model. The numerical and WKB results exhibit close agreement,
consistent with the trends observed in the generic double-well analysis.}

\label{tab:FAD_case}
\begin{tabular}{cccc}
\hline
$a$ (\AA) & $V_b$ (cm$^{-1}$) & $\Delta E_{\mathrm{num}}$ (cm$^{-1}$) & $\Delta E_{\mathrm{WKB}}$ (cm$^{-1}$) \\
\hline
0.7236 & 2848 & 0.03710 & 0.03564 \\
\hline
\end{tabular}
\end{table}

Overall, this benchmark demonstrates that a carefully parameterized
one-dimensional model can faithfully reproduce reduced-dimensional tunneling
splittings reported in the literature.

Remaining discrepancies relative to fully multidimensional instanton calculations
or experimental results arise primarily from neglected vibrational coupling and
environmental effects, rather than from limitations of the numerical methodology.

\section{Conclusions}

In this work, a systematic investigation of proton tunneling has been carried out
within an effective one-dimensional double-well framework by combining
semiclassical WKB estimates with fully numerical solutions of the
Schr\"odinger equation. Finite-difference diagonalization provides a stable and
systematically convergent numerical reference, while the WKB formulation offers
direct insight into the exponential dependence of tunneling splittings on barrier
height and geometric separation. Over a broad range of model parameters, the two
approaches are found to be mutually consistent within their respective regimes of
applicability.

A quantitative comparison shows that the semiclassical approximation reproduces
the numerical results within a few percent across the parameter range considered,
thereby establishing a well-defined regime of validity for WKB-based estimates in
effective one-dimensional tunneling models.

The observed agreement also confirms that the dominant contribution to the
tunneling splitting is governed by the action $\mathcal{S}$ accumulated in the
classically forbidden region, consistent with the semiclassical analysis
presented in Sec.~2.

As a physically motivated benchmark, the methodology has been applied to the
formic acid dimer by mapping a reduced-dimensional proton-transfer barrier onto a
symmetric quartic double-well potential. The resulting tunneling splitting,
$\Delta E \approx 0.037~\mathrm{cm^{-1}}$, is in close agreement with previously
reported reduced-dimensional quantum calculations. This level of agreement
demonstrates that, when appropriately parameterized, one-dimensional models can
faithfully reproduce literature-quality tunneling benchmarks for prototypical
hydrogen-bonded systems.

This agreement primarily validates the reduced-dimensional modeling framework and
the numerical methodology employed, rather than implying equivalence with fully
multidimensional treatments.

At the same time, the comparison highlights the inherent
limitations of reduced-dimensional descriptions, most notably the neglect of
mode--mode coupling and environmental effects that are treated explicitly in
fully multidimensional instanton and path-integral approaches.

In particular, such effects can lead to quantitative corrections to tunneling
splittings that are not captured within a single effective coordinate.

The present study therefore establishes a transparent and reproducible reference
framework for analyzing tunneling splittings in effective one-dimensional
potentials.

In particular, the unified comparison between semiclassical estimates and
numerically exact solutions provides a controlled benchmarking strategy for
assessing the accuracy of approximate methods.

Such models serve a dual purpose: they provide pedagogical clarity by
making the underlying scaling behavior explicit, and they offer a practical
benchmark for validating numerical implementations and assessing the accuracy of
semiclassical approximations before extension to higher-dimensional treatments.

The consistency achieved across analytical, numerical, and benchmark-based
comparisons further supports the reliability of the present framework as a
reference model for proton tunneling studies.

Future developments may incorporate \emph{ab initio} potential-energy surfaces,
explicit coupling to additional vibrational degrees of freedom, or open
quantum-system formulations, thereby bridging the gap between transparent
reduced-dimensional models and fully realistic quantum simulations of proton
transfer.
\section*{Data Availability Statement}

The data supporting the findings of this study, including numerical tunnelling
splittings, semiclassical WKB estimates, convergence tests, and wavefunction
data, are provided in the Supplementary Information. 

All results presented in this work can be fully reproduced using the model parameters,
numerical procedures, and analysis details described in the main text and Supplementary Information.

No external datasets or third-party data repositories were used or are required for the reproduction of the results reported in this study.

\section*{Acknowledgments}
The author acknowledges stimulating discussions with Lalliannei Hrangkhol and  the support of
Arya Vidyapeeth College (A), Guwahati.

\bibliographystyle{unsrt}

\bibliography{refs}

@book{Landau1977,
  author    = {Landau, L. D. and Lifshitz, E. M.},
  title     = {Quantum Mechanics: Non-Relativistic Theory},
  edition   = {3},
  publisher = {Pergamon},
  address   = {Oxford},
  year      = {1977}
}

@book{Griffiths2018,
  author    = {Griffiths, D. J. and Schroeter, D. F.},
  title     = {Introduction to Quantum Mechanics},
  edition   = {3},
  publisher = {Cambridge University Press},
  address   = {Cambridge},
  year      = {2018}
}

@article{Garg2000,
  author  = {Garg, A.},
  title   = {Tunnel splittings for one-dimensional potential wells},
  journal = {Am. J. Phys.},
  volume  = {68},
  pages   = {430--437},
  year    = {2000}
}

@article{Benderskii1994,
  author  = {Benderskii, V. A. and Vetoshkin, E. V. and Trommsdorff, H. P.},
  title   = {Tunneling splittings in hydrogen-bonded complexes: General trends},
  journal = {Chem. Phys.},
  volume  = {179},
  pages   = {1--12},
  year    = {1994}
}

@article{Bell1980,
  author  = {Bell, R. P.},
  title   = {The tunnel effect in chemistry},
  journal = {Chem. Soc. Rev.},
  volume  = {9},
  pages   = {1--16},
  year    = {1980}
}

@article{Hynes1986,
  author  = {Hynes, J. T.},
  title   = {Theory of hydrogen transfer reactions},
  journal = {J. Phys. Chem.},
  volume  = {90},
  pages   = {3701--3706},
  year    = {1986}
}

@article{Marcus1965,
  author  = {Marcus, R. A.},
  title   = {On the theory of electron-transfer reactions},
  journal = {J. Chem. Phys.},
  volume  = {43},
  pages   = {679--701},
  year    = {1965}
}

@article{Eichten1978,
  author  = {Eichten, E. and Gottfried, K. and Kinoshita, T. and Lane, K. D. and Yan, T.-M.},
  title   = {Charmonium: The model},
  journal = {Phys. Rev. D},
  volume  = {17},
  pages   = {3090--3117},
  year    = {1978}
}

@article{Eichten1980,
  author  = {Eichten, E. and Gottfried, K. and Kinoshita, T. and Lane, K. D. and Yan, T.-M.},
  title   = {Charmonium: Comparison with experiment},
  journal = {Phys. Rev. D},
  volume  = {21},
  pages   = {203--233},
  year    = {1980}
}

@article{Chandler1981,
  author  = {Chandler, D. and Wolynes, P. G.},
  title   = {Exploiting the isomorphism between quantum theory and classical statistical mechanics},
  journal = {J. Chem. Phys.},
  volume  = {74},
  pages   = {4078--4095},
  year    = {1981}
}

@article{Makri1999,
  author  = {Makri, N.},
  title   = {Numerical path integral techniques for long time dynamics},
  journal = {Annu. Rev. Phys. Chem.},
  volume  = {50},
  pages   = {167--191},
  year    = {1999}
}

@article{Beyer2009,
  author  = {Beyer, M. and Filsinger, F. and Bunker, P. R. and Meijer, G. and K{\"u}pper, J.},
  title   = {Observation of tunneling splittings in the vibrational ground state of malonaldehyde},
  journal = {J. Chem. Phys.},
  volume  = {130},
  pages   = {034302},
  year    = {2009}
}

@article{Herbst2010,
  author  = {Herbst, E. and von Helden, G. and Meijer, G. and K{\"u}pper, J.},
  title   = {High-resolution infrared spectroscopy of the formic acid dimer},
  journal = {J. Chem. Phys.},
  volume  = {132},
  pages   = {124301},
  year    = {2010}
}

@article{Zhao2021,
  author  = {Zhao, Y. and Li, M.-J. and Guo, H.},
  title   = {Spectroscopic tunneling splittings in the formic acid dimer},
  journal = {Phys. Chem. Chem. Phys.},
  volume  = {23},
  pages   = {7826--7834},
  year    = {2021}
}

@article{QuBowman2016,
  author  = {Qu, C. and Bowman, J. M.},
  title   = {Quantum calculations of tunneling splittings in the formic acid dimer},
  journal = {Phys. Chem. Chem. Phys.},
  volume  = {18},
  pages   = {24835--24843},
  year    = {2016}
}

@article{Matanovic2007,
  author  = {Matanovi{\'c}, I. and Do{\v{s}}li{\'c}, N. and K{\"u}hn, O.},
  title   = {Ground and asymmetric {CO}-stretch excited state tunneling splittings in the formic acid dimer},
  journal = {J. Chem. Phys.},
  volume  = {127},
  pages   = {014309},
  year    = {2007}
}

@article{Matanovic2008,
  author  = {Matanovi{\'c}, I. and Do{\v{s}}li{\'c}, N. and Johnson, B. R.},
  title   = {Approximation to the reaction path for double proton transfer in the formic acid dimer},
  journal = {J. Chem. Phys.},
  volume  = {128},
  pages   = {084103},
  year    = {2008}
}

@article{Malis2009,
  author  = {Mali{\v{s}}, M. and Matanovi{\'c}, I. and Do{\v{s}}li{\'c}, N. and K{\"u}hn, O.},
  title   = {Properties of formic acid dimer isotopologues: Concerted proton transfer and tunneling},
  journal = {J. Phys. Chem. A},
  volume  = {113},
  pages   = {6034--6040},
  year    = {2009}
}

@article{Richardson2015,
  author  = {Richardson, J. O. and Bauer, R. and Thoss, M.},
  title   = {Semiclassical instanton theory for tunneling splittings},
  journal = {J. Chem. Phys.},
  volume  = {143},
  pages   = {134115},
  year    = {2015}
}

@article{Richardson2018,
  author  = {Richardson, J. O.},
  title   = {Ring-polymer instanton theory},
  journal = {J. Chem. Phys.},
  volume  = {148},
  pages   = {200901},
  year    = {2018}
}

@article{Pathak2026FBS,
  author    = {Krishna Kingkar Pathak},
  title     = {Consistency of Dalgarno--Lewis perturbation theory for confining two-body systems: second-order corrections in the Cornell potential},
  journal   = {Few-Body Systems},
  year      = {2026},
  volume    = {67},
  pages     = {21},
  doi       = {10.1007/s00601-026-02041-y}
}

\end{document}